\documentclass[reprint,superscriptaddress,amsmath,amssymb,prb]{revtex4-2}

\usepackage{graphicx}
\usepackage{bm}
\usepackage{hyperref}
\usepackage[usenames,dvipsnames]{xcolor}
\usepackage{listings}

\definecolor{mygreen}{rgb}{0,0.6,0}
\definecolor{mygray}{rgb}{0.5,0.5,0.5}
\definecolor{mymauve}{rgb}{0.58,0,0.82}

\begin{document}
\preprint{APS/123-QED}

\title{Strong correlation effects observed by an ANN-MFT encoder trained on \texorpdfstring{$\alpha$-RuCl$_3$}{alpha-RuCl3} high magnetic field data}

\author{Michael J. Lawler}
\email{mlawler@binghamton.edu}
\affiliation{Dept. of Physics, Applied Physics, and Astronomy, Binghamton University, Binghamton, NY 13902}
\affiliation{Dept. of Physics, Cornell University, Ithaca, NY 14853}
\affiliation{Dept. of Physics, Harvard University, Cambridge, MA 02138}
\author{Kimberly A. Modic}
\affiliation{
Institute of Science and Technology Austria,
Am Campus 1, 3400 Klosterneuburg, Austria}
\author{B. J. Ramshaw}
\affiliation{Dept. of Physics, Cornell University, Ithaca, NY 14853}

\date{\today}

\begin{abstract}
$\alpha$-RuCl3 is a magnetic insulator exhibiting quantum spin liquid phases possibly found in the Kitaev honeycomb model. Much of the effort towards determining Hamiltonian parameters has focused on low magnetic field ordered phases. We study this problem in the high magnetic field limit where mean-field theory is better justified. We do so by machine-learning model parameters from over 200,000 low dimensional data points that include magnetization, torque, and torsion data. Our machine, an artificial neural network-mean-field theory (ANN-MFT) encoder, maps thermodynamic conditions (temperature and field vector) to model parameters via a fully connected time-reversal covariant (equivariant) neural network and then predicts observable values using mean-field theory. To train the machine, we use PyTorch to enable backpropagation through mean-field theory with a pure PyTorch implementation of the Newton-Raphson method. The results at $20$ K and $34.5$ T are consistent with other parameter inference studies in the literature at low magnetic field but strikingly have magnitudes that scale with temperature from 1.3 K up to 80 K in the $34.5-60$ T range
. We conclude that the data presents physics beyond the scope of the mean-field theory and that strong interactions dominate the physics of $\alpha$-RuCl$_3$ up to field strengths of at least 60 Tesla. 
\end{abstract}

\maketitle

\section{\label{sec:introduction}Introduction}
$\alpha$-RuCl$_3$ is a frustrated magnet with unexplained properties. Early experiments observed its insulating character\cite{binotto1971optical} and that quasi-one dimensional $\beta$-RuCl$_3$ orders antiferromagnetically at 600K while the stacked honeycomb structure of $\alpha$-RuCl$_3$ antiferromagnetically orders at 13 K\cite{fletcher1967x}. Experiments in 2015 then recognized that the comparatively low antiferromagnetic transition temperature in $\alpha$-RuCl$_3$ results from frustration, despite the large Curie-Weiss temperature of order 150 K\cite{majumder2015anisotropic,sears2020ferromagnetic} (although the precise size of this number is now in debate\cite{li2021modified}).   In 2017, Neutron scattering then found a continuum of spin excitations at high energy \cite{banerjee2017neutron} with a star-like pattern near the gamma point. Further studies reveal for $H_{ab} > 6T$ and/or $H_{c} > 30 T$, the antiferromagnetic order melts\cite{sears2017phase,baek2017evidence}, a continuum of excitations emerges\cite{wang2017magnetic}, a thermal Hall conductivity appears quantized\cite{kasahara2018majorana} between $6T \lesssim H_{ab} \lesssim 8 T$, scaling emerges in thermodynamic observables\cite{modic2021scale} over a wide range of magnetic fields below 80 Kelvin, and an \emph{in-plane} field generates quantum oscillations\cite{czajka2021oscillations}. How is it that one material can behave in this way?

The simultaneous explanation of these properties is difficult, despite the simplicity of the setting: a van der Waals-coupled honeycomb lattice of spins\cite{burch2018magnetism} that can be studied both in bulk---a kind of ``magnetic graphite''---and as isolated layers---a ``magnetic graphene''. One reason it is hard, is because frustration arises unexpectedly. Since the honeycomb lattice is bipartite, a nearest neighbor Heisenberg model predicts Néel ordering and no frustration is expected. It is therefore the spin-orbit coupling that generates frustration. But this frustration is somehow not only classical, as is understood in Heisenberg magnets in the large-$S$ limit
, but also quantum mechanical---computational studies find small terms normally neglected in spin models dramatically change the phase diagram\cite{lee2020magnetic} and are necessary to stabilize the observed antiferromagnetic order. Another reason an explanation is hard is the initial indications that $\alpha$-RuCl$_3$ is simply described by the Kitaev model has not survived closer scrutiny. Symmetry allows for other terms in the Hamiltonian placing the problem into a more general spin-orbit coupled model space\cite{rau2014generic,katukuri2014kitaev}. Finally, the experiments themselves do not seem to agree with each other. Quantum oscillations from an in-plane field\cite{czajka2021oscillations} clearly evident below 3 Kelvin suggest a quantum spin liquid phase with complex-Fermionic excitations forming a three-dimensional Fermi surface while a half-integer quantized thermal hall effect observed between 3 and 5 Kelvin suggest Majorana fermion edge modes forming a topological phase with gapped fermionic excitations in the bulk\cite{kasahara2018majorana}. 

One strategy to resolve these issues is to infer the Hamiltonian model parameters to enable computational methods to aid in the interpretation of the data. Beginning with the Kitaev model to capture high energy features in neutron scattering, it was argued that ``modest amounts of additional neighbor correlation or simple perturbations based on mean-field approaches'' reproduce the star shaped signal near the Gamma point\cite{banerjee2017neutron}. These fits, either with an antiferromagnetic Kitaev term\cite{banerjee2017neutron} or with ferromagnetic Kitaev coupling, via a parton theory, not not quantitatively reproduce the star pattern\cite{knolle2018dynamics}. But the sign of the Kitaev term is intimately connected to the direction of the ordering moments\cite{chaloupka2016magnetic}. Though hard to determine with neutron scattering\cite{cao2016low}, resonant elastic x-ray scattering (REXS) unambiguously show the Kitaev term is ferromagnetic, a conclusion consistent with fits to many other experiments\cite{sears2020ferromagnetic}. Taken together, these results show the parameter inference problem itself is hard, perhaps only resolved in this region of the phase diagram by a quantum dynamics simulation capable of uniting inelastic and elastic neutron scattering. 

This confusion motivates a second strategy, inferring parameters at large magnetic field. Such a strategy was successful in a quantum spin ice material\cite{thompson2017quasiparticle}]. Furthermore, there is published high field magnetization\cite{johnson2015monoclinic}, torque\cite{modic2018chiral}, and torsion\cite{modic2021scale} data, and even magnetization data in pulsed fields up to 100 T\cite{zhou2022intermediate} data on $\alpha$-RuCl$_3$. Qualitative fits to these datasets\cite{yadav2016kitaev,riedl2019sawtooth,zhou2022intermediate}
are consistent with ferromagnetic Kitaev term. But these fits did not explore the entire parameter space, instead choosing parameters similar to other studies and focused on employing powerful computational methods to compare theory to experiments. So they did not quantify the uncertainty in their fits or take advantage of the simplicity of the high field limit.

In this paper, we combine mean-field theory and artificial neural networks (MFT-ANN) to solve the parameter inference problem at high field. By using an encoding scheme
, we combine magnetization, torque, and torsion data sets into one data set with over 200,000 low-dimensional data points
. The ANN constructs a smooth map from the thermodynamic state $(T,\vec H)$ and data category $C$ of a data point to the Hamiltonian model parameters from which a MFT estimates the experimental observable corresponding to $C$. A trained ANN then shows a variation of these parameters over the data set, a variation that may have physical implications as well as providing a degree of uncertainty quantification. There is no guarantee that training the ANN multiple times will produce the same state-to-parameter map. Instead, the interpretation of the result is reliable when the MFT is able to fit the data well, even if two qualitatively different maps produce equally good fits. Namely, like MFT can have multiple saddle point solutions, data fitting via an MFT-ANN also has this property. Our trained MFT-ANN shows the scaling behavior of thermodynamic observables---previously observed in the intermediate field regime\cite{modic2021scale}---is reflected in the ANN, \emph{with the Kitaev and Gamma couplings scaling with temperature}. Hence, this scaling is a strong correlation effect not captured by MFT and the mapping shows it persists up to 60 T. 

\section{An ANN-MFT encoder for learning model parameters}
\begin{figure}
    \centering
    \includegraphics[width=\columnwidth]{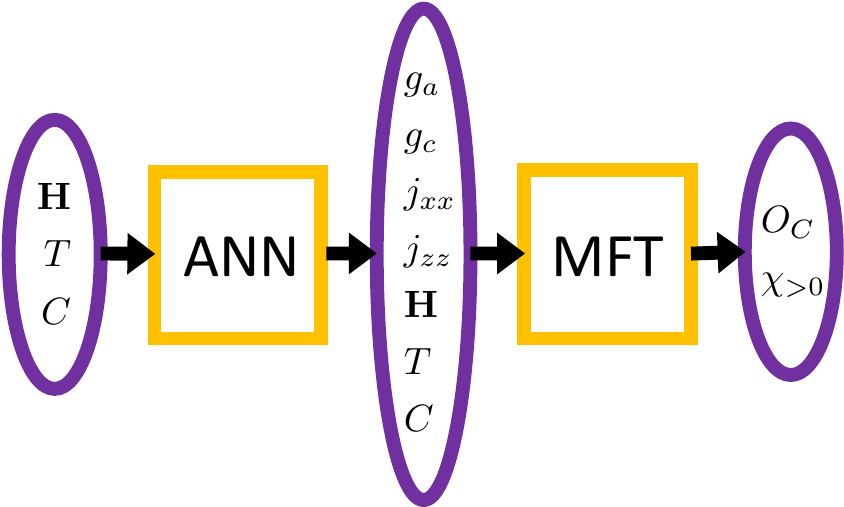}
    \includegraphics[width=0.7\columnwidth]{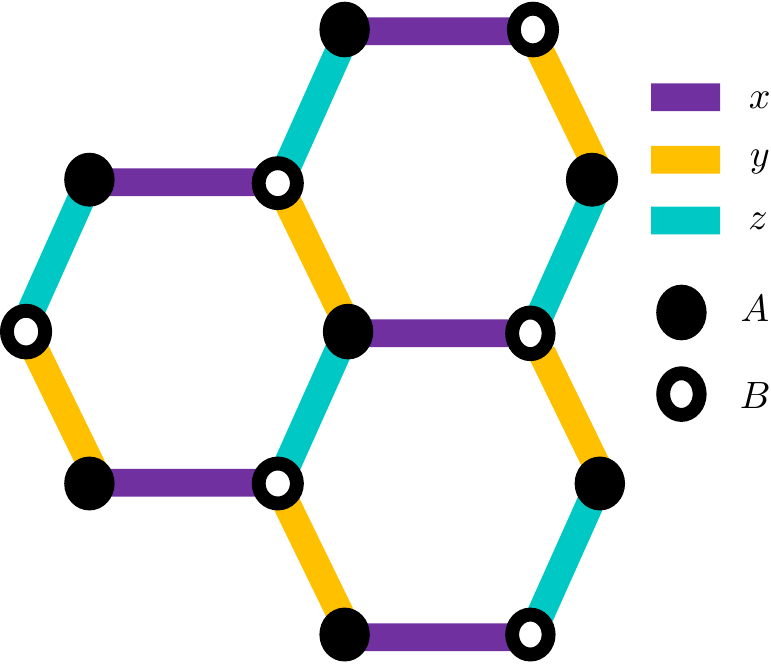}
    \caption{Combining a physics model with a neural network for parameter inference. {\bf a} a flow chart describing the ANN-MFT encoder. Here the thermodynamic state of a condensed matter system, taken to be magnetic field and temperature in this paper, together with an observable category label $C$ is sent through an ANN that encodes this data into parameters of a physics model. A backpropagatable mean-field theory then computes the observable $O_C$ for that model.  {\bf b} The honeycomb-Kitaev model geometry that forms the foundation for a study of $\alpha$-RuCl$_3$. These models live on the honeycomb lattice involving two types of sites, A sites denoted by an open circle, and B sites denoted by a filled circle, and three types of bonds denoted by orange ($x$-bond), purple ($y$-bond), and turquoise ($z$-bond). At each site lives a spin ${\bf S}$ that is anisotropically coupled via g-factors ${\bf g}$ to a magnetic field ${\bf H}$, and on each bond lives an exchange coupling matrix ${\bf J}$ that is different on each of the three types of bonds.}
    \label{fig:model}
\end{figure}

A theory/model of condensed matter physics can be thought of as a decoder. It gives us a map from a set of parameters to a physical system. There are many machine learning models that could be useful for science if we can think of a physics theory as a layer in a machine learning model that allows us to backpropagate through it. Specifically, we need to view the MFT as computing an observable $O$ that is a function of the model parameters $O({\boldsymbol\theta})$ and be able to take derivatives $\partial O/\partial\theta_n$. This will allow the training of a neural network that predicts the model parameters $\theta({\bf W},{\bf b})$ given its weights ${\bf W}$ and biases${\bf b}$ 
. Hence, a backpropagatable MFT layer would solve the problem of adding domain knowledge to a data science algorithm. 

In this paper, we will show how to achieve backpropagation through a mean-field theory(MFT) layer and use it together with an artificial neural network (ANN) encoding layer forming an ANN-MFT encoder as shown in Fig. \ref{fig:model}, to learn the parameters of RuCl$_3$.

In a sufficiently large magnetic field, spins collectively precess about the magnetic field direction and a mean-field ${\bf S}_i \to {\bf m}$ develops. Lowering the magnetic field, antiferromagnetic spin exchange interactions begin to fight this tendancy, and a staggered Neél component to the mean-field ${\bf m}\to{\bf m}_i$ could develop, producing a different mean field on the A and B sublattices. Lowering further, the ordering could become even more complex and/or novel states not captured by mean field theory could arise. 

With this in mind, we begin with the $K$-$\Gamma$ model, believed to capture the physics of RuCl$_3$, defined as
\begin{align}\label{eq:model}
    H_{K-\Gamma}\! &=\! \frac{1}{2}\sum_{\langle ij\rangle} J_{i\alpha,j\beta}S_{i}^\alpha S_{j}^\beta -\!\mu_B \vec H\!\cdot\!{\bf g}\!\cdot\!\sum_i {\bf S}_i\\
    &=\! \sum_{\langle ij\rangle}
    \left[KS_i^\gamma S_j^\gamma\! +\! \Gamma (S_i^\alpha S_j^\beta\!+\!S_i^\beta S_j^\alpha)\right]\!\\\nonumber
    &\quad -\!\mu_B g_a \vec H_\perp\cdot\sum_i {\bf S}^\perp_{i} - \mu_B g_c H_z\sum_i S_i^z
\end{align}
where spin operators are dimensionless with ${\bf S} = \frac{1}{2}(\sigma^x,\sigma^y,\sigma^z)$, $\sigma^\alpha$ the Pauli matrices, and in the second line, we choose the convenient ``theory basis'' to express the couplings $K$ and $\Gamma$ where $\alpha$, $\beta$, $\gamma$ take on the values $(x,y,z)$ if $\langle ij\rangle$ is a ``$z$''--bond, $(z,x,y)$ if $\langle ij\rangle$ is a ``$y$''--bond, and, $(y,z,x)$ if $\langle ij\rangle$ is a ``$x$''--bond, as described in Fig. \ref{fig:model}, and we choose the convenient ``experimental basis'' $(a,b,c)$ to express the g-factors that relate the coupling of the spins to the external magneteic field $\vec H$ in terms of the unit cell geometry, setting $g_a = g_b$ due to assumed rotational invariance in the ab-plane . Since we are interested in thermodynamic data, we then construct the partition function that is the generating function of all thermodynamic observables:
\begin{equation}
    Z = \text{Tr}\exp\bigg(-\frac{1}{2}\sum_{ij\alpha\beta}j_{i\alpha,j\beta}S_i^\alpha S_j^\beta + {\bf h}\cdot\sum_i{\bf S}_i\bigg)
\end{equation}
where $j_{i\alpha,j\beta} = J_{i\alpha,j\beta}/k_BT$ and ${\bf h} = \mu_B {\bf H}\cdot{\bf g}/k_BT$. We write the model parameters in this unusual way because now they are easier to work with in a machine learning context as they are dimensionless parameters of order 1
. In this way, we have set up a model that given $K/k_BT$, $\Gamma/k_BT$, $g_a$, and $g_c$, we can predict any thermodynamic observable provided sufficient computational resources.

The exact evaluation of experimental observables, such as the magnetization $M_\alpha = -k_BT\frac{\partial}{\partial H_\alpha} \ln Z$, is hard for systems with more than 16 spins. So we next make the physically-reasonable mean-field approximation, reasonable as discussed above when there is a large magnetic field. It makes the replacement
\begin{equation}
    S_i^\alpha S_j^\beta =  m_i^\alpha S_j^\beta + S_i^\alpha m_j^\beta - m_i^\alpha m_j^\beta 
\end{equation}
where ${\bf m}_i$ are the mean fields, the mean value of the spin operators $S_i$ on a given site. We have found stopping the complexity of the mean-field theory at the staggered component, i.e. defining just two vectors ${\bf m}_A$ and ${\bf m}_B$ placed in a periodic-in-the-unit-cell arrangement, with sites $i\in A$ belonging to the A sublattice and $i\in B$ belonging to the B sublattice, as shown in Fig. \ref{fig:model}, is sufficient to produce stable solutions to the mean-field equations for typical values of the model parameters. These equations at a finite temperature $T$ are
\begin{equation}\label{eq:mfteqns}
    {\bf m}_A = \frac{1}{|A|}\sum_{i\in A}\langle {\bf S}_i\rangle = \frac{1}{2}\frac{{\bf h}^{eff}_A}{|{\bf h}^{eff}_A|}\text{Tanh}(|{\bf h}^{eff}_A|/2)
\end{equation}
and similarly for ${\bf m}_B$, where ${\bf h}^{\text{eff}}_A = {\bf h} - 3\bar{\boldsymbol \jmath}\cdot{\bf m}_B$ is the dimensionless effective field felt by spins on the A sublattice due to a combination of the external magnetic field and the couplings to the three surrounding spins on the B sublattice. Here $\bar{\boldsymbol\jmath} = {\tt diag}(j_{xx},j_{xx},j_{zz})$ is the average coupling over the three bonds in the unit cell with $j_{xx} = (K-\Gamma)/3k_BT$ and $j_{zz} = (K+2\Gamma)/3k_BT$. Solving these mean-field equations then fixes the values of ${\bf m}_A$ and ${\bf m}_B$ and enables us to compute any observable within the mean-field approximation. 

To solve the parameter inference problem using an ANN-MFT encoder, we need to solve the mean field equations in a way that allows us to backpropagate from a calculated observable like the magnetization, through the model parameters to the weights and biases of the ANN. Defining the ANN as a trainable map $f(x;{\bf W},{\bf b})$ from a point $x\in X$ in a thermodynamic data set $X$ with $x=({\bf H},T)$ to the model parameters $\theta = (g_a,g_c,K,\Gamma)$, with weights and biases ${\bf W}$, ${\bf b}$, and the magnetization computed within the mean-field theory ${\bf M} = {\bf M}(\theta,x)$, we see by the chain rule
\begin{equation}
    \frac{\partial {\bf M}}{\partial W_a} = \frac{\partial{\bf M}}{\partial \theta_m}\frac{\partial\theta_m}{\partial W_a} = 
    \frac{\partial{\bf M}}{\partial \theta_m}\frac{\partial f_m}{\partial W_a},
\end{equation}
where we used $\theta_m = f_m(x;{\bf W},{\bf b})$, that the key challenge is to be able to compute quantities like $\frac{\partial{\bf M}}{\partial K}$ and $\frac{\partial{\bf M}}{\partial g_a}$.

\begin{lstlisting}[float,
caption=Native pytorch code for Newton-Raphson method (also known as Newton's method) adapted from a stackoverflow.com question\cite{stackoverflow}.,
label=lst:newton
]
def newtons_method(function, initial,
    iterations=100, tol=torch.finfo().eps):
  if not(initial.%{\color{NavyBlue}r\,e\,q\,u\,i\,r\,e\,s\,\_\,g\,r\,a\,d}%):
    initial.%{\color{NavyBlue}r\,e\,q\,u\,i\,r\,e\,s\,\_\,g\,r\,a\,d}% = True
  for i in range(iterations): 
    previous = initial.clone()
    value = function(initial)
    value.backward(torch.ones_like(value),
                   retain_graph=True)
    with torch.no_grad():
      initial -= (value / initial.grad)
      initial.grad.zero_()

    if (initial - previous).abs().max() < \
            torch.tensor(tol):
      return initial
  return initial
\end{lstlisting}

The solution we take to accomplish this is numeric differentiation, which is used via automatic differentiation in modern machine learning packages to compute derivatives like $\frac{\partial f_m}{\partial W_a}$ when training the neural network in stochastic gradient descent. As such, we implemented this solution in pytorch with a native implementation of the Newton-Raphston algorithm as shown in Listing \ref{lst:newton}. The key to implementing such a method is to use native pytorch tensors with {\color{NavyBlue}\verb|requires_grad|} enabled. For this to work consistently, it is helpful to use the decorator {\color{Magenta}\verb|@torch|}\verb|.|{\color{NavyBlue}\verb|enable_grad|}\verb|()| above functions that call \lstinline{newtons_method}. Hence, using a native pytorch implementation of the numerical solution to mean-field equations, one can directly use physics models in this approximation as layers within a larger machine learning model connecting experimental data to theoretical predictions.

In addition to backpropagation, a key benefit of using pytorch to implement the solution to the mean field equations is the ability to solve them for different model parameters in parallel on a gpu. We solve them in batches of 1000, a number we found to be efficient while preserving the stochastic property of the gradient descent used to train the ANN. We have been able to solve them in parallel with 100,000 model parameters on a desktop with an Nvidia Titan X GPU, a features that might be valuable in other applications.

\begin{lstlisting}[float,
caption=Code that calculates the susceptibility efficiently.,
label=lst:chi]
  A = Id + 3.0*torch.bmm(chi0,j)
  unstable_chi = torch.zeros(A.shape[0])
  chi = torch.%{\footnotesize\color{NavyBlue}z\,e\,r\,o\,s\_l\,i\,k\,e}%(chi0)
  for i in range(len(A)):
    try:
      chi[i]=torch.linalg.solve(A[i],chi0[i])
    except RuntimeError:
      print("Singular A matrix found!")
      chi[i] = 1.0E6*Id[i]
    try:
      test = torch.linalg.cholesky(chi[i])
    except RuntimeError:
      unstable_chi[i,0] = 1.0
\end{lstlisting}

In the traditional setting, mean-field equations are solved for a particular choice of model parameters and so the mean-fields themselves are chosen to be those which produce stable solutions to these equations. In the setting of this paper, however, we need stable mean-field solutions for all model parameter possibilities, which is not easily achieved. We overcome this challenge by computing the susceptiblity $\boldsymbol\chi$ at the mean-field solution and using this calculation to check if a stable solution was obtained, flagging those that are found to be unstable. Naively, this is done by computing (see Appendix \ref{app:susceptibility} for derivation)
\begin{equation}
    \boldsymbol\chi = ({\bf I} + 3\boldsymbol\chi_0\cdot{\bf j})^{-1}\boldsymbol\chi_0
\end{equation}
There are two ways this calculation can fail. The first is if the matrix ${\bf A} = {\bf I}+3\boldsymbol\chi_0\cdot{\bf j}$ is itself singular. This happens at a phase transition where $\chi\to\infty$ and is not a sign that an unstable solution was obtained. The second is if the matrix $\chi$ is not positive. Such a situation implies the mean-field solution is thermodynamically unstable. We have found the code presented in Listing \ref{lst:chi} was able to perform these checks efficiently. It runs in series on each solution of the mean-field equations previously computed in parallel and achieves efficient checking by a) using \lstinline{torch.linalg.solve} to solve the system of equations ${\bf A}\cdot\boldsymbol\chi = \boldsymbol\chi_0$ for $\boldsymbol\chi$ and use \lstinline{torch.linalg.cholesky} to check via failure of this algorithm for positive-definitness of $\boldsymbol\chi$.

\begin{figure}[t]
    \centering
    \includegraphics[width=\columnwidth]{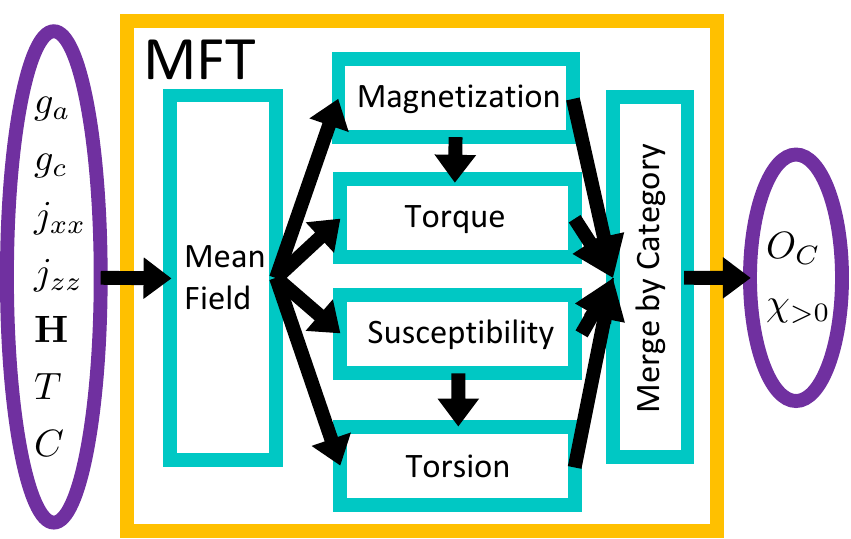}
    \caption{Mean field theory implemented for machine learning purposes in two layers. The first solves the mean field equations to determine the values of the mean fields following Listing \ref{lst:newton}. The second computes the observables for the requested category, either magnetization, torsion, or torque in this study, and a flag $\chi_{>0}$ that is 0 if the mean field solution is unstable and 1 if it is stable.}
    \label{fig:mft}
\end{figure}

Combining the above ideas we can construct the MFT layer as a combination of two layers as shown in Fig. \ref{fig:mft}. One layer solves the mean field equations using Listing \ref{lst:newton}, and the other computes the observables for the category of the data and the flag $\chi_{>0}$ denoting whether or not the MFT equations were stable.  

With the MFT upgraded to perform as a layer in a machine learning algorithm, it remains to specify the ANN. In the context of high magnetic field data, it turns out we can preserve the rotational symmetry about the $c$ axis and the time-reversal symmetry of the data in the architecture of the ANN
, a symmetry covariant/equivariant neural network. To build in the rotational invariance about the $c$-axis we choose the magnetic field ${\bf H}$ to always lie in the $ac$-plane. To build in time-reversal symmetry so that sending in $({\bf H},T,C)$ to the ANN will produce the same model parameters $(g_a,g_c,j_{xx},j_{zz})$ in the output as sending in $(-{\bf H},T,C)$, we need to design the neurons more carefully. 

\begin{figure*}[t]
    \centering
    \includegraphics[width=\textwidth]{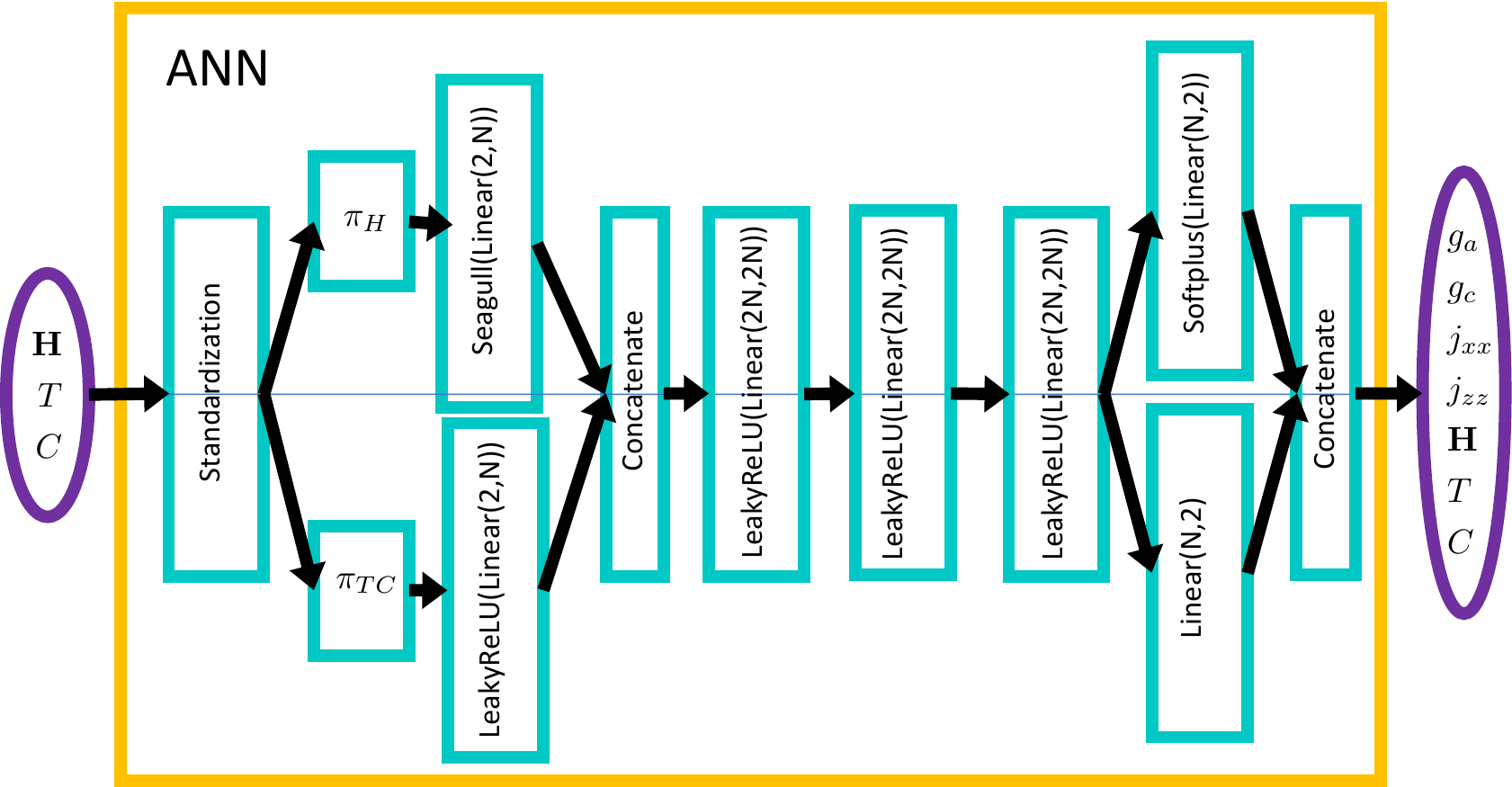}
    \caption{The archecture of the ANN. We choose to build the fully connected neural network that is time-reversal symmetric. We do so by setting the bias to zero and using the symmetric  \lstinline{seagull} activation function in the first layer for magnetic field inputs. Otherwise, the rest of the network is an ordinary fully connected neural network, here using the common \lstinline{LeakyReLU} function with a slope 0.1 for negative values of its input. We also pass the output for parameters $g_a$ and $g_c$ to the \lstinline{softplus} function $\log(1+e^{x})$, a smooth version of \lstinline{ReLU} which guarantees they are positive.}
    \label{fig:ann}
\end{figure*}

We achieve time-reversal symmetry by separating the neurons into two groups, those that act on time-reversal odd variables like ${\bf H}$ and those that act on time-reversal even variables like $T$. Then it is important to preserve this property throughout the network, including when standardizing the data. In general, a neuron acts on data via $\sigma({\bf w}\cdot{\bf x} + {\bf b})$ where ${\bf w}$ is a weight matrix, ${\bf x}$ a data vector, ${\bf b}$ a bias vector, and $\sigma(...)$ is a non-linear function that separately acts on each of the components of the vector in its argument. For a time reversal odd variable, we need to demand $\sigma({\bf w}\cdot(-{\bf x})+{\bf b}) = -\sigma({\bf w}\cdot{\bf x}+{\bf b})$. This requires ${\bf b} = 0$, and $\sigma(-x) = -\sigma(x)$. For the latter, one choice is $\sigma = \text{tanh}$, the hyperbolc tangent function, but we won't need this in our network. For time-reversal even variables, there are no restrictions on $\sigma$, ${\bf w}$, and ${\bf b}$. But since our network outputs only time-reversal even variables, the parameters of the model, we need a neuron that accepts a time-reversal odd variable but outputs a time-reversal even variable. To achieve this, we use the \lstinline{seagull} function $\log(1+x^2)$, previously used in Ref. \onlinecite{gao2020use}. Putting these ideas together, our ANN is presented in \ref{fig:ann}. It immediately converts the time-reversal odd magnetic field variables into time-reversal even ones via the \lstinline{seagull} function and then successive layers are normal with a leakyReLU activation function. We present our ANN in Fig. \ref{fig:ann}. A central benefit of this network is not efficiency, but interpretation for due to the preservation of time-reversal symmetry, angular sweeps of the magnetic field will be strictly periodic in $\theta\to\theta+\pi$ where $\theta$ is the angle of the magnetic field.

\section{Results: scaling and a strong interaction regime}
The MFTANN encoder presented in the previous section was applied to a dataset of 213803 thermodynamic data points taken on $\alpha$-RuCl$_3$ built from a combination of publically available magnetization data at high magnetic fields taken from Ref. \onlinecite{johnson2015monoclinic}, high magnetic field torque data taken from Ref. \onlinecite{modic2018chiral}, and high magnetic field torsion data taken from Ref. \onlinecite{modic2021scale}. We have combined all three thermodynamic data sets into one set by attaching a category label to each measurement observable. This label is converted to a vector via one-hot encoding. Specifically, the in-plane magnetization category was mapped to the vector $(1,0,0,0,0)$, the c-axis magnetization to $(0,1,0,0,0)$, the torque category to $(0,0,1,0,0)$, the torsion field sweep data to $(0,0,0,1,0)$ and the torsion angular sweep data to $(0,0,0,0,1)$. In this way, the ANN can think about these different data types differently due to different weights and biases associated with each in the initial layers. We have made our data set available at zenodo\cite{zenododata}. In addition, we design the MFT layers following \ref{fig:mft} to compute the experimental observables as shown in appendix \ref{app:deriveobs} and combine multiple data types and thereby insert domain knowledge into the machine learning algorithm.

\begin{figure}[t]
    \centering
    \includegraphics[width=\columnwidth]{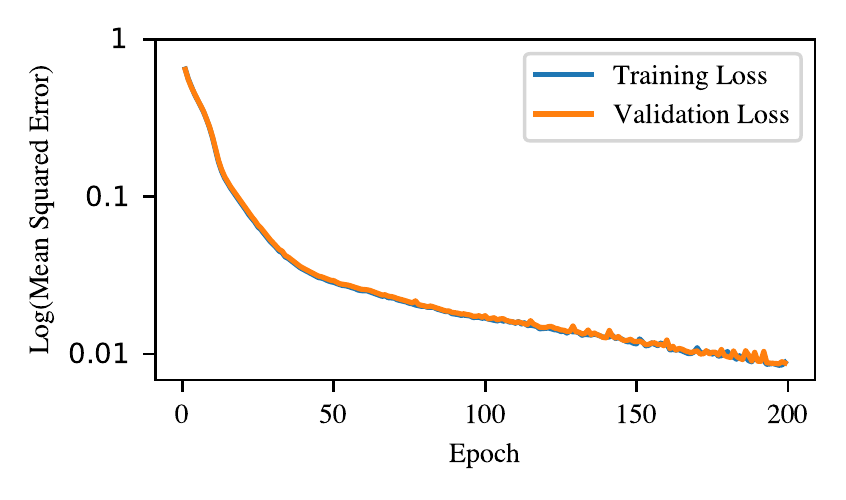}
    \caption{Loss during training. Training loss (blue line) falls slightly below validation loss (orange line) over 200 epochs with a dropout fraction of 0.25 used to control validation loss.}
    \label{fig:training}
\end{figure}

To train the ANN-MFT, we need a loss function that captures the problem. A common choice would be to use the mean-squared error $\frac{1}{|X|}\sum_{(x,y)\in (X,Y)}(\hat y(x)-y)^2$. But it is arbitrary in the sense that it weights all data points equally. It turns out, the data sets are not all equally important for an interpolation scheme occurs when a data set is taken rapidly rendering not all data points as independent measurements. In this case, we can weight each data set separately in the loss function via
\begin{equation}
    L = \sum_n \frac{p_n}{N_n}\sum_{(x,y)\in(X_n,Y_n)}(y(x)-y)^2
\end{equation}
where $N_n$ is the number of data points in data set $n$ and $p_n$ is a normalized weight factor satisfying $p_n>0$, $\sum_np_n=1$. Such a loss function could balance the ANN-MFT so that it weighs each independent data point equally. In this paper, we kept the standard mean-squared-error loss, and its drawbacks, but it would be interesting to explore the alternative loss function in the future. 

An example of the behaviour of the loss during training is shown in Fig. \ref{fig:training}. It shows a rapid decay down to $e^{-5} = 0.0067$ over 200 epochs. Each epoch, the time needed for the machine to see each data point once, took about 10 minutes on our desktop using an Nvidia Titan X GPU. We find that later stages of the training can still have a sizable improvement in the predictions for sparsely populated regions of the data space, regions which are overwhelmed by more densely populated regions at earlier stages of the training. 

In what follows, we will present the predictions of a single trained ANN-MFT model on all the data sets combined as discussed above but then feed this machine individual data sweeps to see what it predicts for those portions of the full data set.

\begin{figure}
    \centering
    \includegraphics[width=\columnwidth]{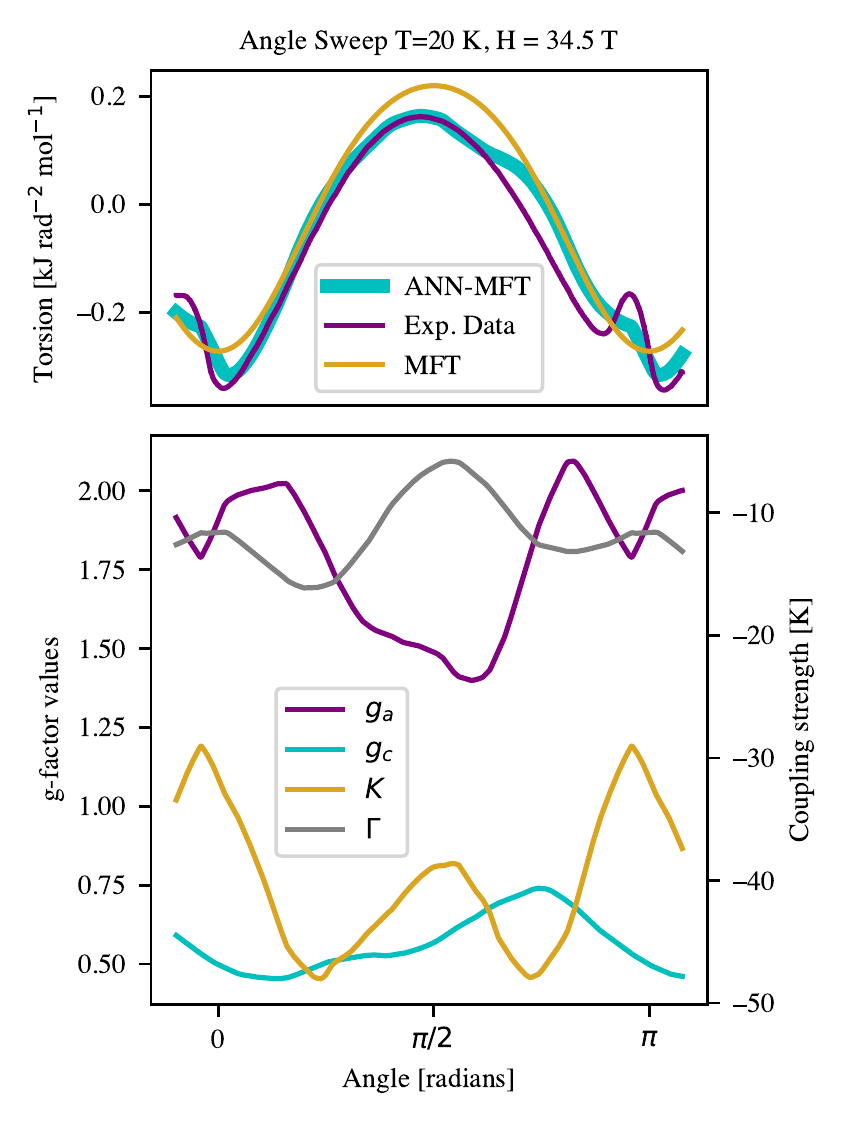}
    \caption{Predictions for the angular sweep torsion data and the couplings used to produce these predictions. (top) Fit to the data showing a good fit throughout much of the angular sweep from $\theta=0$ to $\theta=\pi$, including a decent fit for the MFT predictions with a fixed value of the couplings. (bottom) Couplings used to make the predictions in (top) with the average values of $g_a=1.85$,$g_c=0.54$,$K=-35.26 K$,$\Gamma = -11.80$ chosen for the MFT prediction. Note the discrepancies near the $c$-axis with $\theta=0$.}
    \label{fig:first_torsion}
\end{figure}

The simplest data to interpret are the torsion angle sweeps shown in Figure \ref{fig:first_torsion}. By covering all angles up to a time-reversal symmetry transformation, the ANN-MFT can learn all four parameters $g_a$, $g_c$, $K$, and $\Gamma$. All other data sets fix the angle $\theta$ of the magnetic field to the $c$-axis so, for example, if they are in the $ab$-plane, they would be sensitive to $g_a$ and $j_xx \propto K-\Gamma$ while if they were along the $c$-axis, they would be sensitive to $g_c$ and $j_{zz}$. However, the angle sweeps are at a fixed magnetic field strength of $34.5$ T, so it is unclear if the parameters have saturated to their high-field values. Nevertheless, we find they are roughly constant at all angles and equal to the values $g_a=1.85$, $g_c=0.54$, $K=-35.26$  K,and  $\Gamma = -11.80$ K. These values are consistent with those observed in previous studies, especially note the ferromagnetic nature of the $K$ coupling, that $|K|>|\Gamma|$, and that $g_a>g_c$. 

A second observation about the ANN-MFT predictions for the torsion angle sweep data is the behavior near $\theta=0$. Here we see the data has an unexpected peak which is not present in the mean-field results if we fix the values of the parameters. It is also hard for the ANN-MFT to fit this peak, with its attempt introducing a deviations for angles between $\theta = \pi/2$ to $\theta=\pi$, which was unnecessary between angles $\theta=0$ and $\theta=\pi/2$. We believe this struggle of the ANN-MFT is due to an intrinsically interacting feature in the data not captured by the MFT.

\begin{figure}
    \centering
    \includegraphics[width=\columnwidth]{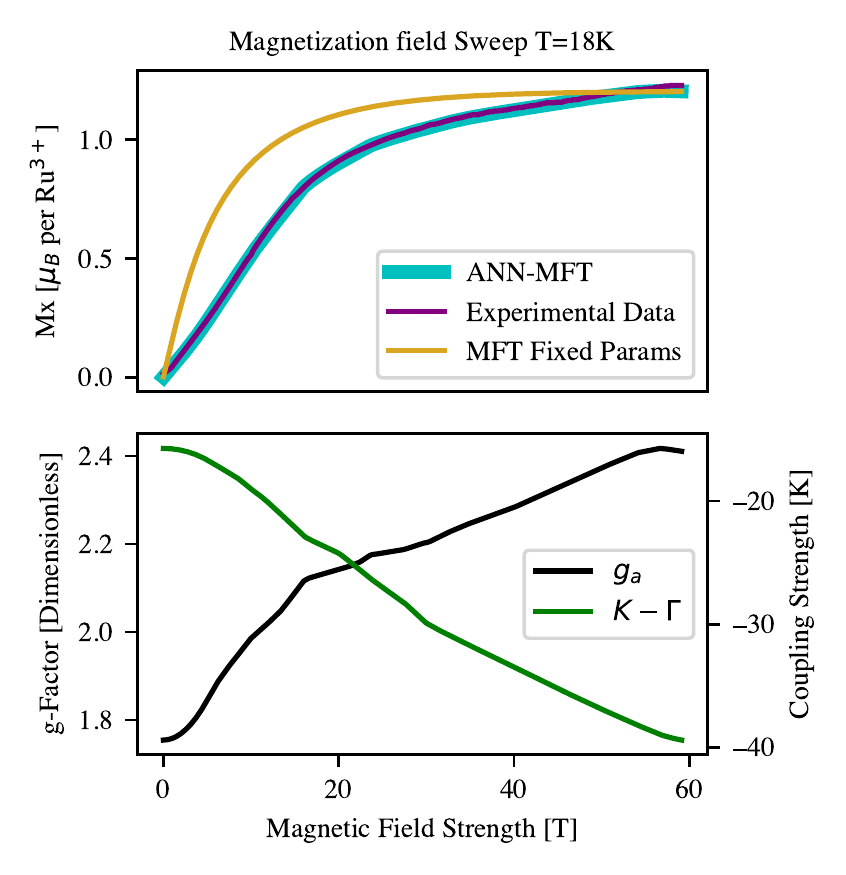}
\caption{Predictions for magnetization data and the couplings used to produce these predictions. {\bf a} predictions for the 18 kelvin, high-field, in-plane magnetization sweep and the prediction of the mean-field theory alone with a fixed choice for the couplings. {\bf b} The variation in the couplings used to make the predictions in {\bf a} during the magnetic field sweep. The fixed choice of couplings for the MFT predictions were taken from the ANN-MFT predictions at 60 Tesla.}
    \label{fig:magnetization}
\end{figure}

The predictions for the magnetization data are shown in Fig. \ref{fig:magnetization}. We see that the $g_a$ g-factor remains relatively stable, ranging between 1.9 and 2.4. The machine cannot accurately predict the $g_c$ value from this in-plane data so we do not present it. The coupling $K-\Gamma$ reaches about -40 Kelvin at 60 Tesla and -30 Kelvin at $H=35 T$, consistent with the torsion angular sweep predictions. 

\begin{figure}
    \centering
    \includegraphics[width=\columnwidth]{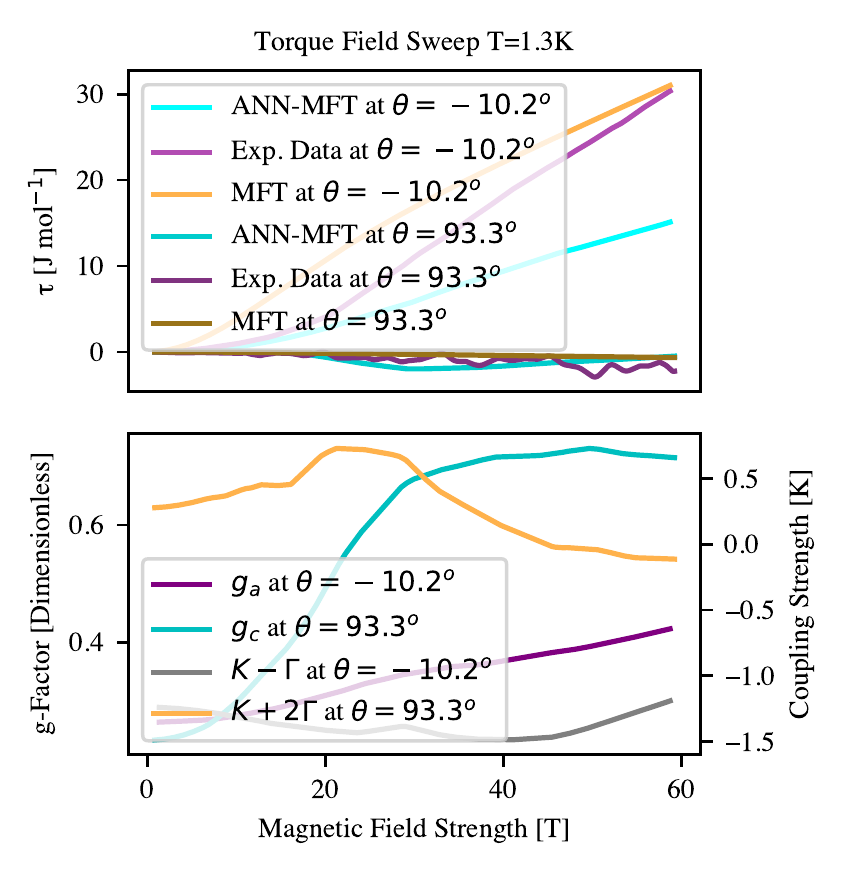}
    \caption{Predictions for the Torque data and the couplings used to produce these predictions along nearly the $a$-axis ($\theta=-10.2^o$) and $c$-axis ($\theta=93.3^o$). (top) Fit to the data showing a relatively poor fit at larger magnetic field strength. We also plot the predictions of the MFT with a fixed set of couplings chosen to match the ANN-MFT predictions at 60 Tesla. (bottom) Couplings used to make the predictions. Torque along the $a$-axis is sensitive to $g_a$ and torque along the $c$-axis is sensitive to $g_c$ so only these are plotted. The Couplings $K$ and $\Gamma$ are much smaller that those found in Fig. \ref{fig:magnetization} but consistent with other data in the data set at low temperatures (here 1.3 Kelvin).}
    \label{fig:torque}
\end{figure}

The torque data shown in Fig. \ref{fig:torque} is harder to interpret than others. Here the ANN-MFT struggled to fit the data so we can trust its predictions less. However, we see the large field predictions at $\theta=-10.2^o$ are $g_a = 0.42$ $K -\Gamma = -1.2 K$, while at $\theta=93.3^o$ are $g_c= 0.71$, $K+2\Gamma=-.11$. Here the machine is predicting $g_a < g_c$, though both are less than 1. solving for $K$ and $\Gamma$ from the two data sweeps, we see it predicts $K=-0.83$ and $\Gamma = 0.36$ and are vastly scaled down in size compared to the predictions of the ANN-MFT on the torsion angle sweep data. 

\begin{figure}
    \centering
    \includegraphics[width=\columnwidth]{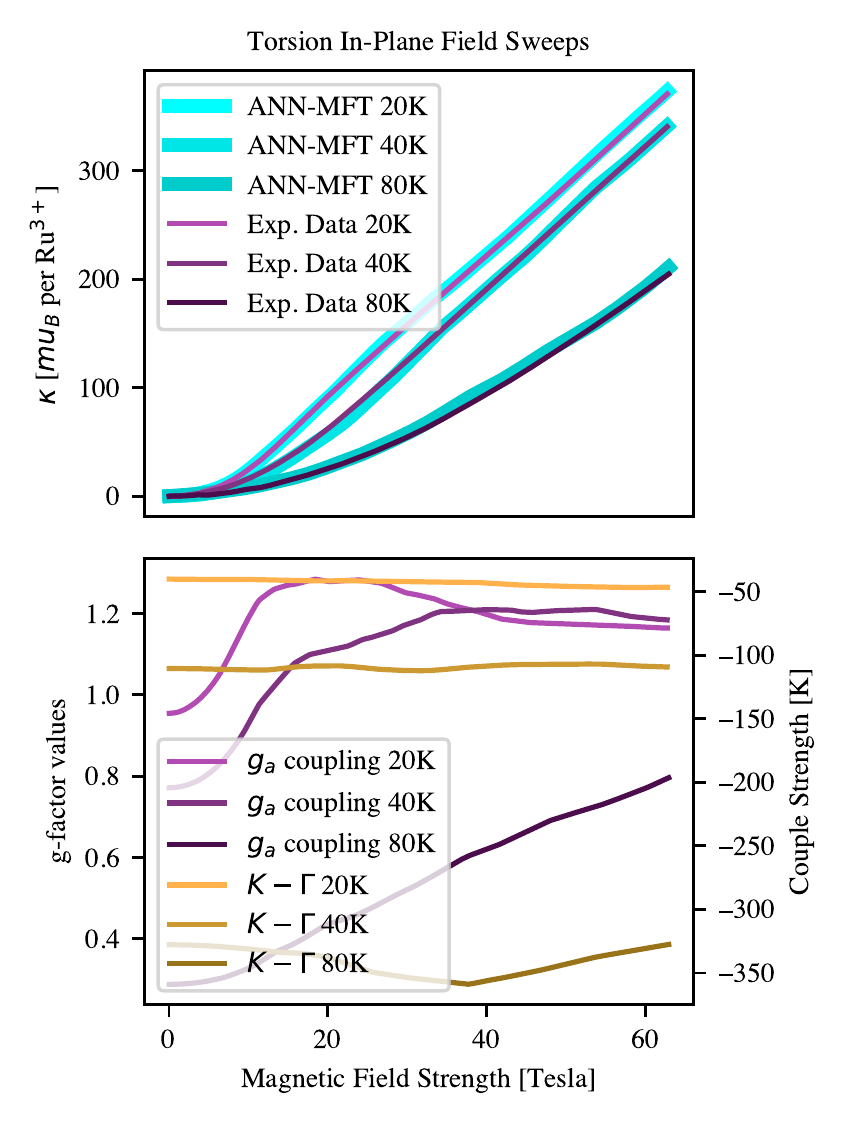}
    \caption{The couplings used to make torsion field sweep predictions at {\bf a} 20K, {\bf b} 40K, and {\bf c} 80K. (top) The ANN-MFT fit to the torsion field sweep data at the three temperatures. (bottom) the $K-\Gamma$ couplings predicted by the ANN-MFT for the three temperatures showing an increase in magnitude for this combination of couplings as a function of temperature approximately doubling between $T=20 K$ and $T=40K$ and again doubling (or more) between $T=40K$ and $T=80K$.}
    \label{fig:torsion}
\end{figure}

The torsion field sweep data presents a further surprise but explains the substantial discrepancy between the torsion angle sweeps and magnetization and the couplings predicted by from the torque data. The couplings predicted from these field sweeps with a $a$-axis magnetic field approximately scale with temperature, doubling in value when the temperature changes from 20K to 40K and again doubling in value when the temperature changes from 40K to 80K as shown in Fig. \ref{fig:first_torsion}. At temperatures above 80K, the doubling stops (not shown). This surprising behavior is consistent with the observed scaling properties of this data set\cite{modic2021scale}. As a result, the torque data now fits into the general picture, it presents scaled down couplings due to the low temperature at which it was taken. If we multiply by the ratio of the temperatures, say for the 40 K torsion field magnitude sweeps, we get ($(K-\Gamma)|_{\text{torque}}*40/1.3 = -36K$, not so far from the $K-\Gamma \approx -100 K$ observed in the torsion at 40K. Hence, the results are roughly consistent across all data sets if one takes into account that the parameters $K$ and $\Gamma$ scale with temperature.

\section{Discussion}
We have introduced an ANN-MFT to study parameter inference with domain knowledge inserted into a machine learning algorithm via a backpropagatable mean-field theory layer. 

There are several weaknesses to this new approach one should be careful of before interpreting the results. 
\begin{enumerate}
    \item Different training can give rise to distinctly different results due to different saddle points of the MFT.
    \item The ANN-MFT may not fit all data points or subsets well.
    \item The ANN-MFT is only as good as the MFT. Is a more general MFT warranted?
    \item What does it mean when we do not discover roughly consistent model parameter values across all data sets?
\end{enumerate}

Let us first address point 1. Through several trials at training the ANN-MFT, it is possible to get wildly different results, perhaps at a cost of a higher loss and poorer fit to the data. These results still fit some data well, but in a different regime of parameters. We interpret this different regime as a saddle point of the MFT. Unlike neural networks that have the magical property that different solutions for the weights behave similarly, the MFT has no such magic. It can have distinct saddle points that correspond to metastable states. It would be interesting in the future to add the energy for each data point to the loss function and either guarantee the results are the lowest energy saddle point solution or use such a modified loss function to isolate metastable states. What we present here is the most common and best trained machines but we do not know if the solution we find is a metastable or stable state. 

For point 2, we emphasize that while the ANN-MFT may not fit all data points well, whenever it does, we can trust the results. If even just a local fit to a portion of the data is good, it implies there is a mapping between thermodynamic state and model parameters that agrees with experiment. 

For points 3 and 4, one possibility is that one should study a more general model, to address point 3, and hope the more general model provides a consistent set of parameters across all data points, to address point 4. Presumably one can generalize the MFT of this paper well beyond the four parameters we study. Introducing the complete spin exchange matrix $J_{i\alpha,j\beta}$ between not only nearest neighbors but also next nearest neighbors and even third neighbors seems quite possible given the efficiency of the algorithm we have presented. Perhaps generalizing the MFT in this way will yield consistent parameter values across all data sets. But this doesn't seem likely in this case since a scaling with temperature does not seem to be readily captured by further neighbor couplings. 

In addition to the weaknesses, there are many ways to improve the simple ANN parameter inference method used in this manuscript. One is to upgrade the ANN encoder to behave like a variational autoencoder that encodes not directly to the parameters but instead to a probability distribution from which a parameter sample is drawn. While such a probabilistic approach would seem to introduce more noise into the machine, variational autoencoders are actually more efficient than ordinary autoencoders. Such an upgrade, in addition to efficientcy, would then provide error bars on the prediction of the parameter values. 

Despite the weaknesses and simplicity of the ANN-MFT machine we have used to study $\alpha$-RuCL$_3$ high magnetic field data, we find it striking that a scaling with temperature is the predominant observation across the entire data set from three different experiments. A naive explanation for scaling is that at these large magnetic fields, the system becomes approximately non-interacting. By scaling, $\kappa/T = f(\mu_BH/k_BT,K/k_BT, \Gamma/k_BT)$. If we can set $(K,\gamma)/k_BT\approx 0$, then we obtain $\kappa/T = f(\mu_BH/k_BT)$, a scaling function. But our observations are that $K$ and $\Gamma$ are proportional to temperature. Hence, $\kappa/T$ scales but we are never in the non-interacting regime. We further notice near fields pointing along the $c$-axis, that anomalies appear in the data that the ANN-MFT cannot fit accurately (see torsion angle sweep results). This is strong evidence that the temperature regime $1.3 K$ to $80 K$ is captured by an interacting theory beyond the Curie-Weiss mean-field theory approximation. 

Unlike the data set as a whole, data at temperatures between 20 and 80 K saturated at large magnetic field. It would be nice to understand this from a perturbation theory perspective. In the limit $|{\bf h}|\gg |\bar{\boldsymbol\jmath}|$, owing to a gap in the spectrum, the system is adiabatically connected to a non-interacting paramagnet. However, the limit we find our selves in this manuscript, as it appears within MFT say from the torsion angle sweeps, is $|\bar{\boldsymbol\jmath}| = max(|K-\Gamma|/3k_BT,|K+2\Gamma|/3k_BT) = 0.98$ when $0.6 \lesssim|{\bf h}|\lesssim 1.7$ so that both $|{\bf h}|$ and $|\bar{\boldsymbol\jmath}|$ are of the same order or magnitude. These values do not to appear to appreciably change with temperature from 1.3 Kelvin to 80 Kelvin and only above this temperature do we begin to see a departure. This suggests this entire regime is not easily captured by a perturbation theory and so there is no controlled approximation to study it.

Spin liquids like $\alpha$-RuCl$_3$ are hard to study experimentally, yet an ANN-MFT or similar machine could prove to be a powerful tool. It is striking that from bulk data we obtain parameters that are consistent with information rich techniques like neutron scattering and x-ray scattering. We believe ANN-MFTs and related machine learning techniques will enable a powerful relationship between experimental data and model parameters, especially in the common situation of many spin liquid candidate materials where neutron scattering and X-ray scattering data are unavailable.

\begin{acknowledgments}
This material is based upon work supported by the National Science Foundation under Grant No. OAC-1940260.
\end{acknowledgments}

\appendix

\section{Derivation of mean-field observables}
\label{app:deriveobs}
To derive observables within the mean field theory, we begin with the mean field partition function with mean fields ${\bf m}_A$ and ${\bf m}_B$ satisfying Eq. \ref{eq:mfteqns}
\begin{equation}
    Z_{MF}\! =\! \text{Tr} \exp\!\bigg\{\!-{\bf h}^{eff}_A\!\cdot\!\sum_{i\in A}{\bf S}_i - {\bf h}^{eff}_B\!\cdot\!\sum_{i\in B}{\bf S}_i+3\beta N_B{\bf m}_A{\bf \bar J}{\bf m}_B\!\bigg\}
\end{equation}
where $N_B$ is the number of bonds, ${\bf \bar J} = \frac{1}{NB}\sum_{\langle ij\rangle}{\bf J}_{ij}$ is the average exchange matrix, ${\bf h}^{\text{eff}}_A = {\bf h} - 3{\bf j}\cdot{\bf m}_B$ is the effective magnetic fields felt by spins on the A sublattice due to their interactions with spins on the B sublattice and similarly for the ${\bf h}^{eff}_B$ with A and B swapped. We can evaluate this partition function directly and obtain
\begin{multline}
    Z_{MF} = 2^{|A|+|B|}e^{N_B{\bf m}_A{\bf \bar J}{\bf m}_B}\\
    \times\cosh^{|A|}\left(\left|{\bf h}^{eff}_A/2\right|\right)
    \cosh^{|B|}\left(\left|{\bf h}^{eff}_B/2\right|\right)
\end{multline}
From this partition function we can obtain all the observables we need to compare with the experimental data. Specifically, we can compute the magnetization, magnetic susceptibility, magnetic torque, and magnetic torsion observables, as shown in the next few sections.

\subsection{Magnetization}
To compute magnetization, we use the definition
\begin{equation}
    M^\alpha = -\frac{\partial F_{MF}}{\partial H^\alpha}
\end{equation}
where $F_{MF} = -k_BT\ln Z_{MF}$ and $H^\alpha$ is a component of the external magnetic field ${\bf H}$. 
Evaluating this expression gives
\begin{multline}
    M^\alpha = k_BT\frac{|A|}{2}\frac{\partial |{\bf h}^{eff}_A|}{\partial H^\alpha}\tanh\left(\left|{\bf h}^{eff}_A/2\right|\right) + A\to B +\\
    \frac{\partial}{\partial H^\alpha}\left(N_B{\bf m}_A{\bf\bar J}{\bf m}_B\right)
\end{multline}
where
\begin{equation}
    \frac{\partial |{\bf h}^{eff}_A|}{\partial H^\alpha} = {\bf\hat h}^{eff}_A\cdot \frac{\partial {\bf h}^{eff}_A}{\partial H^\alpha}.
\end{equation}
Recognizing that ${\bf m}_A = {\bf\hat h}^{eff}_A\tanh\left(\left|{\bf h}^{eff}_A/2\right|\right)$ we see this simplifies to
\begin{equation}
    M^\alpha = k_BT|A|{\bf m}_A\cdot\frac{\partial {\bf h}^{eff}_A}{\partial H^\alpha} + A\to B + \frac{\partial}{\partial H^\alpha}\left(N_B{\bf m}_A{\bf\bar J}{\bf m}_B\right).
\end{equation}
Using
\begin{equation}
    \frac{\partial {\bf h}^{eff}_A}{\partial H^\alpha} = \frac{\mu_B}{k_BT}{\bf e}^\alpha\cdot{\bf g} - 3{\bf j}\frac{\partial{\bf m}_B}{\partial H^\alpha}
\end{equation}
with ${\bf h} = \mu_B {\bf H}\cdot{\bf g}/k_BT$ then obtain
\begin{multline}
    M^\alpha = \mu_B(|A|{\bf m}_A+|B|{\bf m}_B)\cdot {\bf g}\cdot{\bf e}^\alpha\\
    -3|A|{\bf m}_A{\bf \bar J}\frac{\partial{\bf m}_B}{\partial H^\alpha} - 3|B|{\bf m}_B{\bf \bar J}\frac{\partial{\bf m}_A}{\partial H^\alpha} + \frac{\partial}{\partial H^\alpha}\left(N_B{\bf m}_A{\bf\bar J}{\bf m}_B\right)
\end{multline}
where we used ${\bf j} = {\bf \bar J}/k_B T$. For periodic boundary conditions, $|A| = |B|=N_u$ and $N_B = 3 N_u$ and we see that the quadratic inn ${\bf m}_A$, ${\bf m}_B$ terms cancel leaving us with
\begin{equation}\label{eq:mag}
    M^\alpha = N\mu_B{\bf\bar m}\cdot {\bf g}\cdot{\bf e}^\alpha
\end{equation}
We see then this result could have been obtained another way. The cancellation is precisely what is required to allow us to first compute $M^\alpha$ exactly and then perform the mean-field approximation. 

Numerically, for an Avogadro's number of atoms, we can express this as $M^{a,b} = (\mu_B/k_B)g_a R(m^{a,b}_A + m^{a,b}_B)/2$ and $M^c = (\mu_B/k_B)g_c R(m^c_A+m^c_B)/2$ where $\mu_B/k_B = 0.6714 [K/T]$ and $R=8.314 [J/K]$. Or for a data set expressed in units of per Bohm magneton per spin we would use $M^{a,b} = g_a(m^{a,b}_A + m^{a,b}_B)/2$ and $M^c = g_c(m^{c}_A + m^{c}_B)/2$, as is the case of the data studied in this manuscript.

\subsection{Magnetic Torque}
To compute the torque ${\boldsymbol \tau}$, we can use the definition
\begin{equation}
    {\boldsymbol\tau} = {\bf M}\times{\bf H}
\end{equation}
and take the component in the direction of increasing azimuthal angle ${\boldsymbol\phi}$ to obtain:
\begin{equation}
    \tau_\phi = \hat{\boldsymbol\phi}\cdot({\bf M}\times{\bf H}) = {\bf M}\cdot({\bf H}\times\hat{\boldsymbol\phi}) = -H{\bf M}\cdot\hat{\boldsymbol\theta}
\end{equation}
We recognize we can compute this directly from the free energy
\begin{equation}
    \tau_\phi \equiv \frac{\partial F}{\partial\theta} = \frac{\partial H^\alpha}{\partial\theta}\frac{\partial F}{\partial H^\alpha} = -H\hat{\boldsymbol\theta}\cdot{\bf M}
\end{equation}
Here and throughout this paper we choose the spherical polar coordinate system ${\bf H} = H(\cos\phi\sin\theta,\sin\phi\sin\theta,\cos\theta)$ with $\hat{\boldsymbol\phi}$ and $\hat{\boldsymbol\theta}$ the directions of increasing $\phi$ and $\theta$ respectively. If we were to follow Ref. \cite{modic2018chiral} and parameterize the magnetic field \emph{from the ab plane} as ${\bf H} = H'(\cos\phi'\cos\theta',\sin\phi'\cos\theta',\sin\theta')$ then we would find $\hat{\boldsymbol\theta'} = -\hat{\boldsymbol\theta}$ and $\partial F/\partial\theta' = -\partial F/\partial\theta$. As a result, the torque data presented in this paper differs in the sign convention for the observable $\tau_\phi$. 

A convenient numerical expression for $\tau_\phi$, with an Avogadro's number of atoms and a magnetic field in the $ac$-plane, is obtainable by inserting Eq. \ref{eq:mag} in to our expression for $\tau$ and writing it as
\begin{equation}
   \tau = -RT \frac{d{\bf h}}{d\theta}\cdot\bar{\bf m}
\end{equation}
where $R=8.314[J/K]$ and
\begin{equation}
    \frac{d{\bf h}}{d\theta} = \frac{\mu_B}{k_BT}(g_aH^c,0,-g_cH^a).
\end{equation}
We are free to place ${\bf H}$ in the $ac$-plane, i.e. set $\phi=0$, because the rotation symmetry about the $c$-axis allows us to always choose this unknown parameter. Fixing $\phi=0$ then always produces mean field solutions with ${\bf m}_A$ and ${\bf m}_B$ also in the $ac$-plane so we choose to ignore the $y$ components in our calculation of observables.

\subsection{Magnetic Susceptibility and Thermodynamic Stability}\label{app:susceptibility}
The magnetic susceptibility is defined as
\begin{equation}
    \chi_{\alpha\beta} = -\frac{\partial^2 F}{\partial H^\alpha\partial H^\beta}
\end{equation}
Technically speaking it is defined in the limit $|{\bf H}|\to0$. However, at any finite magnetic field ${\bf H}$, thermodynamic stability demands
\begin{equation}
    \delta {\bf H}\delta{\bf M} + \delta T\delta S + \delta\mu\delta N \geq 0
\end{equation}
so that using $\delta M^\alpha = \partial M^\alpha/\partial H^\beta\delta H^\beta=\chi^{\alpha\beta}\delta H^\beta$ we see that ${\boldsymbol\chi}\geq0$ both at finite and the limit of zero magnetic field. Hence, the observable ${\boldsymbol\chi}$ is a useful quantity at any value of the magnetic field.

It is worthwhile simplifying the computation of the magnetic susceptibility. It is related to the spin susceptibility, via the chain rule
\begin{equation}
    \chi^{\alpha\beta} = -\frac{\partial h^\gamma}{\partial H^\alpha}\frac{\partial^2 F}{\partial h^\gamma\partial h^\delta}\frac{\partial h^\delta}{\partial H^\alpha} = \frac{\mu_B^2}{k_BT} g^{\alpha\gamma}\chi_s^{\gamma\delta}g^{\delta\beta}
\end{equation}
where we used $\partial h^\alpha/\partial H^\beta = \mu_Bg^{\alpha\beta}/k_BT$ and defined the spin susceptibility as $\chi_s^{\alpha\beta} = \frac{\partial^2}{\partial h^\alpha\partial h^\beta}(-\log Z)$. Since it is straightforward to convert from ${\boldsymbol\chi}$ to ${\boldsymbol\chi}_s$, via ${\boldsymbol\chi} = (\mu_B^2/k_BT){\bf g}{\boldsymbol\chi}_s{\bf g}$, we will proceed by focusing on the simpler ${\boldsymbol\chi}_s$.

We can further simplify the problem of computing the magnetic susceptibility. Within the MFT, the spin susceptibility is the derivative  $\chi^{\alpha\beta}_s=\partial \bar m^\alpha/\partial H^\beta$ with $\bar{\bf m} = \frac{1}{2}\sum_\mu{\bf m}_\mu$, where $\mu = A,B$ denotes sublattice (see Eq. \ref{eq:mag}). Hence we need compute $\partial m^\alpha_\mu/\partial H^\alpha$. We can do so by computing the sublattice dependent spin susceptibility $\chi^{\alpha\beta}_{s\mu\nu} = \partial m^\alpha_\mu/\partial H^\beta_\nu$ where we have introduced a hypothetical magnetic field ${\bf H}_A$ and ${\bf H}_B$ that acts separately on the $A$ and $B$ sublattices and this new susceptibility captures the response to a change in the magnetic field in just one of the sublattices. It is related to $\chi_s^{\alpha\beta}$ by the chain rule
\begin{equation}\label{eq:chichi6}
    \chi^{\alpha\beta}_s = \sum_\mu\frac{\partial m^\alpha_\mu}{\partial h^\beta} = \sum_{\mu\nu}\frac{\partial m^\alpha_\mu}{\partial h^\beta_\nu}\bigg|_{{\bf h}_A = {\bf h}_B = {\bf h}}
\end{equation}
where we thought of ${\bf m}_\mu$ as a function of two independent fields ${\bf h}_A$ and ${\bf h}_B$, i.e. ${\bf m}_\mu({\bf h}_A,{\bf h}_B)$ and then took the derivative of ${\bf m}_\mu({\bf h},{\bf h})$ with respect to ${\bf h}$. As a result, we can focus on the sublattice dependent susceptiblity $\chi^{\alpha\beta}_{s\mu\nu}$, a quantity we can compute efficiently as a 6x6 matrix. An additional benefit is this quantity also must satisfy ${\bf \chi}_6\geq 0$ by thermodynamic stability and so fully expresses whether the mean field theory is stable. 

To compute ${\boldsymbol\chi}_6$, it is helpful to work with a 6-component vector notation, ${\bf m}_6={\bf m}_A\oplus{\bf m}_B$, combining the three components of magnetization on the A sublattice and B sublattice. In this language, the mean-field equations define a non-linear map
\begin{equation}
    {\bf m}_6 = {\bf V}({\bf h}_6 - 3\bar{\boldsymbol \jmath}_6\cdot{\bf m}_6)
\end{equation}
where ${\bf h}_6 = {\bf h}\oplus{\bf h}$,  $\bar{\boldsymbol\jmath}_6 = \bar{\bf J}_6/k_BT$ with ${\bf J}_6 = \bar{\bf J}\otimes\sigma_x$, and ${\bf V}$ is the map ${\bf V}({\bf x}_6) = \frac{1}{2}{\bf x}_6\odot W({\bf x}_6)$ where
\begin{equation}
    {\bf W}({\bf x}_6) =
    \begin{pmatrix}
    |{\bf x}_A|^{-1}\tanh(|{\bf x}_A|/2)\\
    |{\bf x}_A|^{-1}\tanh(|{\bf x}_A|/2)\\
    |{\bf x}_A|^{-1}\tanh(|{\bf x}_A|/2)\\
    |{\bf x}_B|^{-1}\tanh(|{\bf x}_B|/2)\\
    |{\bf x}_B|^{-1}\tanh(|{\bf x}_B|/2)\\
    |{\bf x}_B|^{-1}\tanh(|{\bf x}_B|/2)
    \end{pmatrix}
\end{equation}
with $\odot$ denoting the broadcast matrix operation such as ${\bf A}\odot{\bf B} = (A_1B_1,A_2B_2,A_3B_3)$. Hence by the chain rule, we have
\begin{equation}
    \nabla_{h_6}{\bf m}_6 = {\boldsymbol\partial V}\cdot\left({\bf I}_6-3\bar{\boldsymbol \jmath}\nabla_{h_6}{\bf m}_6\right)
\end{equation}
and so the sublattice dependent spin susceptibility ${\boldsymbol\chi}_6 \equiv \nabla_{h_6}{\bf m}_6$ is given by
\begin{equation}\label{eq:chi6}
    {\boldsymbol \chi}_6 = \left({\bf I}_6+3{\boldsymbol\partial V}\bar{\boldsymbol \jmath}_6\right)^{-1}{\boldsymbol\partial\bf V}
\end{equation}
where we recognize ${\boldsymbol\partial\bf V} = {\boldsymbol\chi}_s^0$ is the ``bare'' sublattice spin susceptibility evaluated at the effective magnetic field ${\bf h}_6-3\bar{\boldsymbol\jmath}{\bf m}_6$. Evaluating the derivative, we see it is given by
\begin{multline}
    {\boldsymbol\partial V}({\bf x}_6) = \frac{1}{2}{\tt diag}(W({\bf x}_6)) +\\ \frac{1}{2}\left[({\bf x}_A\otimes{\bf x}_A)T(|{\bf x}_A|)\right]\oplus\left[({\bf x}_B\otimes{\bf x}_B)T(|{\bf x}_B|)\right]
\end{multline}
where $T(x) = \frac{1}{2x^2}(1-\tanh(x/2)^2)-\tanh(x/2)/x^3$. Hence, we have reduced the calculation of the magnetic susceptibility ${\bf\chi}$ to the determination of $\chi_6$ via Eq. \ref{eq:chi6}. 

The implementation of Eq. \ref{eq:chi6} was discussed in the main manuscript and was done robustly via Listing \ref{lst:chi}. A highlight of this calculation was the determination of wether the mean-field theory was thermodynamically stable. This was achieved by checking whether the Cholesky decomposition failed with a \lstinline{try}-\lstinline{catch} statement. In this way, unstable solutions to the mean-field theory could be caught. We chose to deal with such cases by dropping those data points from the loss function. So long as the number of data points that map to unstable MFTs were small, say 1\% or 2\%, we found the overall training of the ANN-MFT worked successfully in that it defines a mapping onto the parameters of the Hamiltonian that accurately fits the data for most data points. 

\subsection{Torsion}
The last observable we need to calculation is torsion $\kappa$. It is related to magnetizataion via 
\begin{equation}
    \kappa \equiv \frac{\partial^2 F}{\partial\theta^2} = \frac{\partial\tau_\phi}{\partial\theta}
\end{equation}
By chain rule, we can write this as
\begin{equation}
    \kappa = \frac{\partial}{\partial\theta}\bigg(\frac{\partial H^\alpha}{\partial\theta}\frac{\partial F}{\partial H^\alpha}\bigg) = -\frac{\partial^2 {\bf H}}{\partial\theta^2}\cdot{\bf M} - \frac{\partial {\bf H}}{\partial\theta}\cdot\frac{\partial{\bf M}}{\partial\theta}
\end{equation}
Recognizing that $\partial^2{\bf H}/\partial\theta^2 = -{\bf H}$ and using $\partial{\bf H}/\partial\theta = H\hat{\boldsymbol\theta}$ we obtain
\begin{equation}
    \kappa = {\bf H}\cdot{\bf M} - H\hat{\boldsymbol\theta}\cdot\frac{\partial {\bf M}}{\partial\theta}
\end{equation}

It remains to place $\kappa$ in a numerically convenient form. By using Eq. \ref{eq:mag} and ${\bf h} (\mu_B/k_BT){\bf g}\cdot{\bf H}$, and the chain rule, we can write $\kappa$ as
\begin{equation}
    \kappa = \frac{RT}{2}\left(2{\bf h}\cdot\bar{\bf m} - \frac{d{\bf h}}{d\theta}\cdot{\boldsymbol\chi}_s\frac{d{\bf h}}{d\theta}\right)
\end{equation}
and further in the 6-dimensional vector notation it becomes
\begin{equation}
    \kappa = \frac{RT}{2}\left({\bf h}_6\cdot{\bf m}_6 - \frac{d{\bf h}_6}{d\theta}\cdot{\boldsymbol\chi}_6\frac{d{\bf h}_6}{d\theta}\right)
\end{equation}
where we used Eq. \ref{eq:chichi6} to relate ${\boldsymbol\chi}_s$ and ${\boldsymbol\chi}_6$. 
This is the form we use in our calculations.


%

\end{document}